# Multi-Comm-Core Architecture for Terabit/s Wireless

Farooq Khan
Samsung Electronics
Richardson, Texas, USA

*Abstract* — Wireless communications along with the Internet has been the most transformative technology in the past 50 years. We expect that wireless data growth driven by new mobile applications, need to connect all humankind (not just 1/3) as well as Billions of things to the Internet will require Terabit/s shared links for ground based local area and wide area wireless access, for wireless backhaul as well as access via unmanned aerial vehicles (UAVs) and satellites. We present a new scalable radio architecture that we refer to multi-comm-core (MCC) to enable low-cost ultra-high speed wireless communications using both traditional and millimeter wave spectrum.

Keywords: Wireless, low-power, radio architecture, Terabit/s, cellular, Wi-Fi, backhaul, UAV, satellite communications.

## I. INTRODUCTION

Mobile connectivity has transformed daily life across the globe becoming one of the most dramatic game-changing technologies the world has ever seen. As more people connect to the Internet, increasingly chat to friends and family, watch videos on the move, and listen to streamed music on their mobile devices, mobile data traffic continues to grow at unprecedented rates. Actually, this surge in demand is following what we can informally call an *omnify* principle. Here, *omnify* stands for *Order of Magnitude Increase every Five Years*. Which means, demand for data increases 10 times every 5 years and will continue to increase at this pace with expected 1,000 times increase in the next 15 years. This increase in demand is similar to the memory and computing power growth following Moore's law which offered a million fold more memory capacity and the processing power in the last 30 years. For wireless communications, it is more appropriate to measure advances in 5 years and 10 years timeframe as a new generation wireless technology is introduced every 10 years and a major upgrade on each generation follows 5 years afterwards as shown in Figure 1. The total global mobile traffic already surpassed 1 Exabyte/ month mark in 2013 and is projected to grow 10 fold exceeding 10 Exabytes/ month within 5 years in 2018 [1]. With this trend, in 2028 global mobile traffic will exceed 1 Zettabyte/ month which is equivalent to 200 Gigabytes/month for 5 Billion users worldwide. Wi-Fi offload accounts for almost as much traffic as carried on mobile networks and also follows a similar growth trend with expected 1 Zettabyte/ month Wi-Fi offload traffic in 2028. We also expect peak wireless data rates to follow *omnify* principle, increasing from 1 Mb/s in year 2000 with 3G to around 10 Gb/s with 5G in 2020 and finally 1 Tb/s with 6G in 2030 offering million times increase in 30 years as depicted in Figure 1. Others have also highlighted the need and drivers for Tb/s data rates [2]. The peak data rates for Wi-Fi follow a similar trend with a few years lead. This means wireless data is catching up with the memory, storage and computing capabilities that are already available to deal with these massive amounts of data.

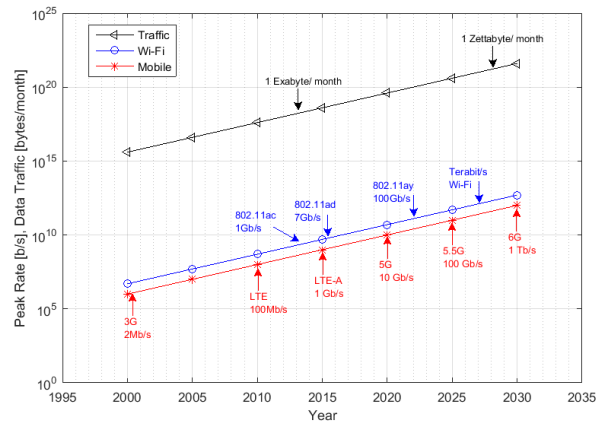

**Figure 1 Wireless data follows an omnify principle**

Traditionally, all wireless communications, with the exception of point-to-point microwave backhaul links and satellite communications, used a relatively narrow band of the spectrum below 3GHz. This sub-3GHz spectrum has been attractive for Non-Line-of-Sight (NLOS) point-to-multipoint wireless communications due to its favorable propagation characteristics. The large antenna aperture size at these frequencies enables broadcasting large amount of power which allows signals to travel longer distances as well as bend around and penetrate through obstacles more easily. This way, the sub-3GHz spectrum allowed providing wide area coverage with a small number of base stations or wireless access points (WAPs). However, with mobile data traffic explosion, modern wireless systems face capacity (not coverage) challenges requiring deployment of more and more base stations with smaller coverage area. However, the number of small cells that can be deployed in a geographic area is limited due to the costs involved for acquiring the new site, installing the equipment and provisioning backhaul etc. In theory, to achieve 1,000-fold increase in capacity, the number of cells also needs to be increased by the same factor. Therefore, small cells alone are not expected to meet the capacity required to accommodate orders of magnitude increase in mobile data traffic demand in a cost effective manner.

In order to address the continuously growing wireless capacity challenge, the author and his colleagues pioneered use of higher frequencies above 3GHz referred to as millimeter waves with a potential availability of over 250GHz spectrum for mobile communications [3]-[5]. At millimeter wave frequencies, radio spectrum use is lighter and very wide bandwidths along with a large number of smaller size antennas can be used to provide orders of magnitude increase in capacity needed in the next 15 to 20 years. The smaller size of antennas is enabled by carrier waves that are millimeters long compared to centimeters long waves at currently used lower frequencies. A drawback of mmWaves, however is that they tend to lose more energy than do lower frequencies over long distances, because they are readily absorbed or scattered by gases, rain, and foliage.

The adaptive beamforming technology can overcome these challenges by using an array of smaller size antennas to concentrate radio energy in a narrow, directional beam, thereby increasing gain without increasing transmission power. A prototype of adaptive beamforming using a matchbook-size array of 64 antenna elements connected to custom-built signal-processing components was demonstrated in [6]. By dynamically varying the signal phase at each antenna, this prototype transceiver generates a beam just 10 degrees wide that it can switch rapidly in the desired direction. The base station and mobile radio continually sweep their beams to search for the strongest connection, getting around obstructions by taking advantage of reflections thereby providing Non-line-of-sight (NLOS) communications.

## II. TERABIT WIRELESS

As of 2014, almost two-thirds of the humankind does not have access to the Internet. A major barrier to expanding access to these communities is the cost of providing mobile services. Therefore, our foremost goal is to reduce cost per bit by developing a new wireless architecture and integrated solution that can scale to Terabit/s for ground based local area and wide area wireless access, for wireless backhaul as well as access via unmanned aerial vehicles (UAVs) and satellites as depicted in Figure 2. In many parts of the World where high-speed data communications infrastructure is not available, it is considered more economical to provide high-speed access via satellites, unmanned aerial vehicles (UAVs) or other non-traditional systems. The satellite communications and wireless backhaul markets have been highly fragmented due to use of proprietary systems. One reason for this fragmentation is dependence upon lower frequencies (below 6GHz) for local area and wide area wireless access while wireless backhaul and satellite communications operate at higher millimeter wave spectrum.

However, as discussed earlier, with the vision of providing wireless mobile access in the millimeter wave spectrum, a single wireless technology can be developed for access, backhaul, aerial and space systems eliminating fragmentation and thereby reducing costs of providing wireless services across these diverse platforms. With the use of higher millimeter wave frequencies, back-haul links and access can also share the same spectrum due to highly directional nature of beamformed millimeter wave transmissions. However, for satellite and /or UAV based communications with very wide coverage area, methods to avoid or cancel interference needs to be considered when the same spectrum is shared between the different platforms.

We note that battery-powered handheld mobile devices may not be able to transmit (or even receive) terabit/s data rates due to high power consumption. Therefore, terabit/s access links would need to be shared among multiple of such devices with each device capable to transmit or receive at peak data rates say of the order of tens of Gigabit/s. However, backhaul links between wireless access points (WAPs), between ground station and a satellite, between a ground station and a UAV or link between a satellite and an airplane or a base station and a car, for example, can transmit and receive at peak data rates of Terabit/s or higher. We also note that in some of these cases the distinction between access link and backhaul link becomes blurred. For example, an airplane can provide access service for its passengers who are using battery-powered handheld devices while connected to the satellite using so called sky-haul. A UAV can use either sky-haul link to the satellites or connect to the ground system via a backhaul link.

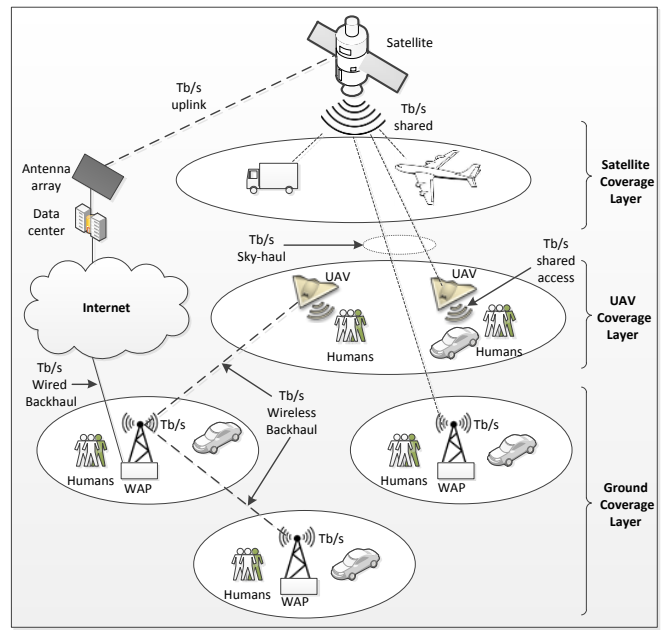

**Figure 2 Terabit/s for ground-based, UAV and satellite wireless links**

We expect that future mobile and Wi-Fi systems will continue to rely on below 6GHz frequencies for important control information transmission and for data communications when higher frequencies signals are not available. This is because radio waves below 6GHz frequencies can better penetrate obstacles, and are less sensitive to non-line-of-sight (NLOS) communications or other impairments such as absorption by foliage, rain, and other particles in the air.

A key benefit of using higher frequencies is wider available bandwidths providing higher data rates and capacity thereby lowering cost per bit. In order to fully utilize the lower and higher frequencies spectrum, we provide a three-layer system design framework as shown in Figure 3. The lower-band-group layer utilize spectrum below 6GHz providing a maximum peak data rate of around 10Gb/s using arrays consisting of tens of antennas. The mid-band-group covers spectrum between 6-56GHz providing peak data rate of 100Gb/s using arrays consisting of hundreds of antennas. In the high-band-group, peak data rates approache Terabit/s with possibility of using thousands of antennas.

A single system design approach can be used for the three layers with each layer potentially having different physical layer parameters to optimize performance. This is because the channel propagation characteristics are expected to be different in the three band groups due to dependence upon frequency. At the same time, we expect that the differences in channel characteristics within a band group small enough to justify a single set of physical layer parameters. For example, from Figure 3, we can see that High-band system needs to account for higher attenuations due to water vapor ($H_2O$) and oxygen ($O_2$) absorption [7].

### A. Mid-band spectrum

The use of lower-band-group spectrum below 6GHz is well understood and therefore we start by exploring the mid-band spectrum that covers 6-56GHz frequencies range. In October 2004, FCC issued a Notice of Inquiry (NOI) to examine use of bands above 24 GHz for mobile radio services [8]. FCC's UK counterpart, Ofcom, also released a Call for Input on "Spectrum above 6 GHz for future mobile communications" [9]. The bands identified below 60 GHz in the FCC NOI providing a total bandwidth of 5.2GHz are summarized in Table 1 and Figure 3. The regulators in other countries are also currently working on identifying millimeter spectrum for wireless communications. We defined mid-band spectrum in the range of 6-56GHz covering a total 50GHz spectrum. Therefore, the FCC NOI only identifies about 10% of the total mid-band spectrum. We expect that in the future, more spectrum in the mid-band can be identified for wireless communications not only in the United States but also globally. Separate spectrum allocations are provided for satellite communications as summarized in [11].

### B. High-band spectrum

Terabit/s wireless would require tens of GHz of spectrum which is practically impossible in the low-band below 6GHz and unlikely in the mid-band due to presence of other services. However, much larger spectrum bandwidths can be made available in the high-band which covers above 56GHz frequencies. In Figure 3 and Table 1, we also summarize the 60/ 70 / 80 GHz bands identified in the FCC NOI. We also list 90GHz and higher bands that are currently designated as Mobile in FCC frequency allocation chart [10] in addition to other services such as passive earth exploration satellite, radio astronomy, passive space research, fixed / mobile satellite, radio navigation, amateur radio and non-communication ISM Part 18 equipment etc. We note that a total 66.6GHz bandwidth can potentially be available in the high-band up to 164GHz to enable Terabit/s wireless communications. However, the use of wireless and mobile communications in these bands needs to be carefully studied to evaluate impact on other services present in these bands. We further remark that with continuous advances in technology, bands above 164 GHz can also become candidates for future wireless communications in the future.

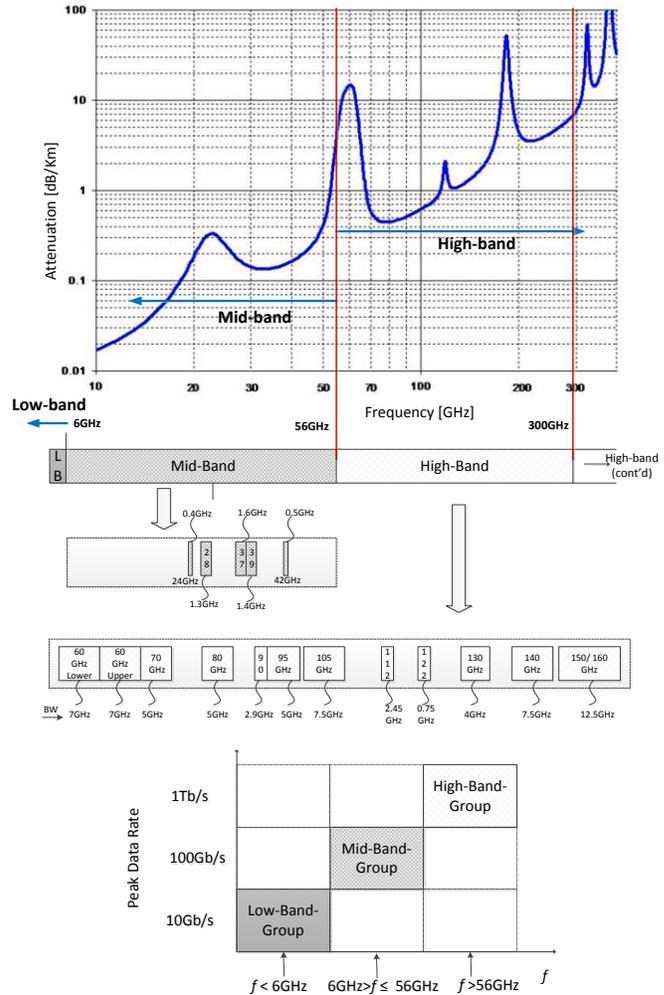

**Figure 3. Mid-band and High-band Millimeter wave spectrum**

**Table 1. Mid-band and high-band spectrum candidates**

| Bands [GHz] | Frequency [GHz] | Bandwidth [GHz] | |
|---|---|---|---|
| 24 GHz Bands | 24.25-24.45 | 0.200 | |
| | 25.05-25.25 | 0.200 | Mid-band spectrum |
| LMDS Band | 27.5-28.35 | 0.850 | |
| | 29.1-29.25 | 0.150 | |
| | 31-31.3 | 0.300 | |
| 39 GHz Band | 38.6-40 | 1.400 | 5.2 GHz |
| 37/42 GHz Bands | 37.0-38.6 | 1.600 | |
| | 42.0-42.5 | 0.500 | |
| 60 GHz | 57-64 | 7.000 | High-band spectrum = 66.6 GHz |
| | 64-71 | 7.000 | |
| 70/80 GHz | 71-76 | 5.000 | |
| | 81-86 | 5.000 | |
| 90 GHz | 92-94 | 2.900 | |
| | 94.1-95.0 | | |
| 95 GHz | 95-100 | 5.000 | |
| 105 GHz | 102-105 | 7.500 | |
| | 105-109.5 | | |
| 112 GHz | 111.8-114.25 | 2.450 | |
| 122 GHz | 122.25-123 | 0.750 | |
| 130 GHz | 130-134 | 4.000 | |
| 140 GHz | 141-148.5 | 7.500 | |
| 150/ 160 GHz | 151.5-155.5 | 12.50 | |
| | 155.5-158.5 | | |
| | 158.5-164 | | |
| >164 GHz | High-Band Future Extensions | | Over 100 GHz spectrum |

### III. MULTI-COMM CORE ARCHITECTURE

Multiple computational cores or multi-core have become the norm ever since IBM introduced the first general-purpose processor called POWER4 in 2001 that featured multiple processing cores on the same CMOS die. The transition to multi-core architecture was driven by unsustainable level of power consumption implied by clock rates in excess of a few GHz. The heat losses and dynamic power of integrated circuits (the power that is consumed when the transistor is switching from an on-state to an off-state) both increase faster as frequencies rise. The switching or dynamic power dissipated by a CMOS chip increases quadratically with voltage as below.

$$P \propto CV^2 f + P_{static}$$

where $C$ is the capacitance being switched per clock cycle, $V$ is the supply voltage, $f$ is the switching frequency, and $P_{static}$ is the power due to static leakage current, which has become more and more accentuated as feature sizes have become smaller and threshold levels lower. The voltage required for stable operation is determined by the frequency at which the circuit is clocked, and can be reduced if the frequency is also reduced. This can yield a significant reduction in power consumption because of quadratic relationship above. Currently, the only practical way to improve the processing performance is to keep the clock rates around 1-2 GHz while adding support for more threads, either in the number of cores, or through multithreading on cores. We expect a similar transition will happen in wireless communications when data rates in the hundreds of Gigabit/s to Terabit/s range would require many GHz of bandwidth.

We present a scalable radio system architecture labeled as Multi-Comm-Core (MCC) shown in Figure 4 to achieve these ultra-high speed data transmissions with reasonable complexity and power consumption. As we go to such high data rates while using very large bandwidths, it is more energy efficient to work in smaller blocks operating at lower clock frequencies and using multiple of them together to achieve the high data rate. We already see this trend in current multi-core CPU designs for microprocessors, graphic cards and mobile phones where exploiting parallelism is a key to attain performance at limited power consumption. With MIMO and carrier aggregation support in current wireless systems, we are already in the era where we are having multiple similar blocks of hardware in the system and we are seeing it pervade in all aspects of the design at the micro and macro levels such as antenna arrays (with multiple identical elements), Analog-to-Digital Converters (with time interleaved sub-ADCs), LDPC decoders (multiple decoders with multiple identical processing elements within each decoder) and so on. The proposed radio architecture also complements the multi-core processor architecture used for the application processor which can implement the MAC, security and higher layers functions.

We can also consider evolution of HW accelerators used in the multi-core application processors to align with the proposed multi-comm-core radio architecture. Moreover, in case of Cloud-RAN or C-RAN implementations, we can centralize and aggregate baseband processing part of the MCC architecture for a large number of distributed radio nodes.

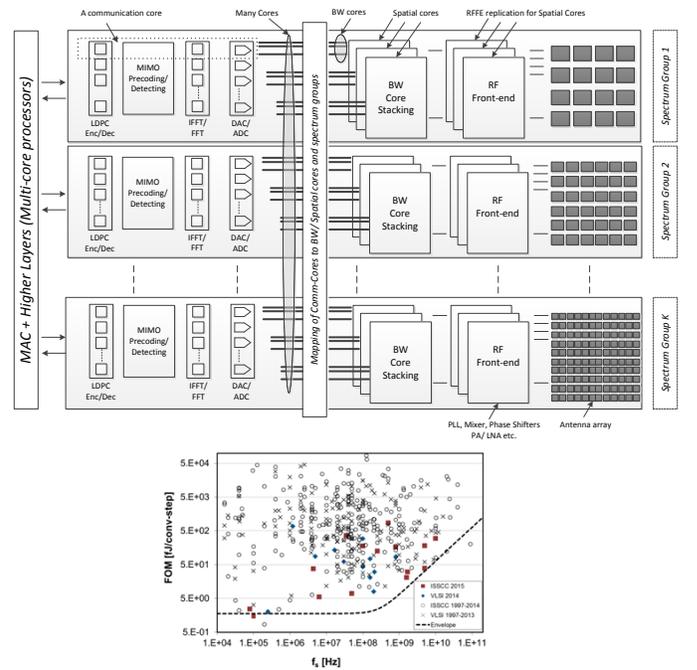

**Figure 4 Multi-Comm-Core (MCC) Architecture and Walden FOM vs speed for ADCs [13]**

In practice, the bandwidths required to support ultra-high speed data transmissions may not be available in a contiguous manner (see Table 1) even in the higher millimeter waves or the RF front-end and the antennas may not support such high bandwidths with required efficiency. Therefore, we expect the system to use a set of RF front-end and antenna arrays covering each spectrum group of frequencies as shown in

Figure 4. Within each spectrum group, multiple bandwidth (BW) cores each supporting 1-2GHz bandwidth will be stacked together. The RF front-end and BW core stacking blocks within each spectrum group are replicated to provide spatial cores (to support multi-beam and/or MIMO capability). Each BW core has its own set of data converters (ADC/DACs), FFT/IFFT, channel coding and other baseband functions and these blocks/functions are replicated across spatial cores. The MIMO/beam processing may need to be performed jointly across the spatial cores for the same BW core.

The proposed radio architecture provides high degree of flexibility to achieve a given data rate through use of various combinations of BW cores, spatial cores and spectrum groups based on HW and spectrum availability. Moreover, it allows scaling down the HW to meet given cost, form factor, power consumption or complexity requirements.

An example of key block that will benefit from multi-comm-core architecture are data converters. A commonly used Figure of merit (FOM) for ADCs referred to as Walden FOM [12] defines energy required per conversion step as below.

$$FOM = \frac{P}{2^{ENOB} \times f_s} \quad [J/conv]$$

where $P$ is the power dissipation, $f_s$ is the Nyquist sampling rate, and ENOB is the effective number of bits defined by the signal-to-noise and-distortion ratio (SNDR).

Figure 4 shows this FOM as function of speed or the Nyquist sampling rate for many ADCs published in the literature. We note that energy required per conversion step increases by orders of magnitudes when going above GHz sampling rates. This makes a case for limiting the sampling rates to around a GHz and using many ADCs, a pair of ADCs for I and Q sampling per comm-core. We note that with the semiconductor process technology scaling, the cut-off point for transitioning to multi-comm-cores may move to higher sampling rates and hence higher bandwidths.

The energy of clock distribution network of integrated circuits scales with the clock frequency and consumes a significant portion of the energy as well [14]. The MCC architecture allows keeping the clock frequencies low thereby providing not only savings in power consumption but also reduction in heat that needs to be dissipated. By limiting the core bandwidth to around 1GHz, MCC architecture also avoids or minimizes beam squint (i.e., changes in the beam steering angle with frequency) issues when true-time delay (TTD) cannot be implemented such as may be the case for baseband beamforming.

The MCC architecture presents several other advantages from system capacity and hardware perspective. For example, the number of bandwidth MCCs allocated for uplink and downlink can be dynamically varied to meet the varying traffic needs while fully utilizing the air-interface and hardware resources. Also, not all MCCs need to be homogeneous meaning supporting the same bandwidth or number of antennas etc. A set of heterogeneous MCCs can take into account different requirements of hardware and spectrum allocation as well as needs of low-power control signals transmission or low-rate data transmission. Moreover, similar to multi-core CPU architecture, we can turn on only the required number of MCCs to support a given data rate and capacity at a given time turning of the other MCCs to save power.

IV. LINK BUDGET ANALYSIS

We carry out the link budget analysis for a Terabit/s system in Table 2. We first calculate the data rate per comm-core for a reference system with 1GHz core bandwidth at 100 GHz frequency and a 200 meters range. We note that with a total of 256 comm-cores, we can reach 1.5 Terabit/s data rate. This can, for example, be achieved with 32 BW cores with total 32 GHz bandwidth and 8 spatial cores or other combinations of BW cores and spatial cores. This means that very large bandwidths and large number of spatial cores (multiple beams) will be required to achieve Terabit/s data rates. With the proposed MCC architecture, the total bandwidth (32GHz in this example) can also be aggregated across low-band, mid-band and high-band spectrum. Another observation is large amount of total transmit power which will be 25.6 Watts for 256 cores just for the power amplifiers only with the assumption of 100mW of power per core. Accounting for power amplifier efficiency and consumption in other RF components and the baseband, we expect the total power consumption at least an order of magnitude higher in the hundreds of Watts range or higher. As discussed earlier, this will limit the feasibility of Terabit/s access data rates in the downlink from a WAP, UAV or satellite to mobile devices. In case of back-haul links between base stations or a link between a satellite and a ground-based gateway, both ends of the link can possibly transmit and receive at Terabit/s data rates. Moreover, this aggregate data rates would need to be shared among multiple mobile devices as just receiving Terabit/s data rates will be way above the power budget available in mobile devices. This can be achieved by sharing the resources in frequency, time and space (multi-user MIMO) among multiple users. With the assumption that a mobile device can transmit over a few cores within its power budget, a reasonable assumption for the data rates from a mobile device then probably is in the range of tens of Gigabits/s.

The trasmit antenna total gain of 32 dB is achieved using an array of size 512 and anenna element gain of 5 dB. We assume a receive antenna array size of 64 elements providing a total gain of 23 dB after accounting for anenna element gain of 5 dB. When antenna arrays are used both at the transmitter and the receiver, higher frequencies will benefit from higher array gain for the same total antenna area due to smaller wavelengths [1][4]. This will help compensate for other losses that tend to increase with increasing frequency. In the link budget calculations, we accounted for total losses of 18 dB which includes 10 dB for NLOS reflection, 3dB RF front-end loss and 5 dB baseband implementation loss. In practice, these losses may be higher and there may be other losses which will reduce achievable data rate, range or both. For satellite based access requiring very large distance links, achieving Terabit/s data rates would require to increase the transmit power and antennas gain to compensate for the higher propagation loss as discussed in [11].

**Table 2. Terabit/s link budget analysis**

| Parameter | Value | Comments |
|---|---|---|
| Transmit Power | 20 dBm | Possibly multiple PAs |
| Transmit Antenna Gain | 32 dBi | Element + array gain |
| Carrier Frequency | 100 GHz | Ref. for calculations |
| Distance | 200 meters | |
| Propagation Loss | 118.42 dB | |
| Other path losses | 10 dB | Some NLOS |
| Tx front end loss | 3 dB | Non-ideal RF |
| Receive Antenna Gain | 23 dB | Element + array gain |
| Received Power | -56.42 dBm | |
| Bandwidth (BW) | 1 GHz | BW / comm-core |
| Thermal Noise PSD | -174 dBm/Hz | |
| Receiver Noise Figure | 5.00 dB | |
| Thermal Noise | -79 dBm | |
| SNR | 22.58 dB | |
| Implementation loss | 5 dB | Non-ideal baseband |
| Spectrum Efficiency (SE) | 5.86 b/s/Hz | |
| **Data rate / comm-core** | **5.86 Gb/s** | SE × BW |
| Number of comm-cores | 256 | BW and spatial cores |
| **Aggregate data rate** | **1.5 Terabit/s** | 256×5.86 Gb/s |

## V. CONCLUSION

We outlined a vision to connect remaining 2/3 of the humankind as well as Billions of things to the Internet via low-cost Terabit/s wireless access. We presented a scalable radio architecture referred to as multi-comm-core (MCC) that can scale to higher frequencies millimeter wave spectrum in addition to using traditional frequencies below 6GHz. With the MCC architecture, number of communication cores and hence the bandwidth and the hardware can be scaled naturally to provide terabit/s wireless for ground based local area access (Wi-Fi), wide area wireless (mobile) access, wireless backhaul as well as access via unmanned aerial vehicles (UAVs) and satellites. A single radio technology operating at both traditional and millimeter wave spectrum can be developed eliminating fragmentation and thereby reducing costs of providing wireless services through these diverse platforms.

ACKNOWLEDGMENT

The author would like to thank his colleagues at SAMSUNG for valuable discussions and feedback.